\documentclass[preprint,review,12pt]{elsarticle}

\usepackage{graphicx}
\usepackage[nodots]{numcompress}
\usepackage{epstopdf}
\biboptions{sort&compress}

\begin{document}
\begin{frontmatter}

\title{EELS study of the epitaxial graphene/Ni(111) and graphene/Au/Ni(111) systems}

\author[FHI,TUDD,SPB]{A. V. Generalov}
\author[FHI,TUDD]{Yu. S. Dedkov\corref{corr}}

\address[FHI]{\mbox{{Fritz-Haber-Institut der Max-Planck-Gesellschaft, 14195 Berlin, Germany}}}
\address[TUDD]{\mbox{{Institut f\"ur Festk\"orperphysik, Technische Universit\"at Dresden, 01062 Dresden, Germany}}}
\address[SPB]{\mbox{{V.A. Fock Institute of Physics, St. Petersburg State University, 198504 St. Petersburg, Russia}}}

\cortext[corr]{Corresponding author. E-mail address: yuriy.dedkov@gmail.com (Yu. S. Dedkov)}

\begin{abstract}
We have performed electron energy-loss spectroscopy (EELS) studies of Ni(111), graphene/Ni(111), and the graphene/Au/Ni(111) intercalation-like system at different primary electron energies. A reduced parabolic dispersion of the $\pi$ plasmon excitation for the graphene/Ni(111) system is observed compared to that for bulk pristine and intercalated graphite and to linear for free graphene, reflecting the strong changes in the electronic structure of graphene on Ni(111) relative to free-standing graphene. We have also found that intercalation of gold underneath a graphene layer on Ni(111) leads to the disappearance of the EELS spectral features which are characteristic of the graphene/Ni(111) interface. At the same time the shift of the $\pi$ plasmon to the lower loss-energies is observed, indicating the transition of initial system of strongly bonded graphene on Ni(111) to a quasi free-standing-like graphene state. 
\end{abstract}


\end{frontmatter}

\section{Introduction}

The electron energy-loss spectroscopy (EELS) in reflection mode with an electron beam of low primary energy is very suitable method for characterization and investigation of true and quasi two-dimensional (2D) systems, due to its high surface sensitivity~\cite{Rocca:1995}. Graphene, the planar sheet of $sp^2$-bonded carbon atoms packed in a honeycomb lattice, can be considered as a good example of such systems. This material has received enormous attention because of its unique electronic band structure and physical properties~\cite{Geim:2007,Geim:2009}. Recent EELS studies of the freestanding graphene layer in the transmission mode~\cite{Eberlein:2008} and epitaxial graphene on SiC(0001) in the reflection mode~\cite{Langer:2009,Lu:2009} have demosntrated a strong sensitivity of these methods to the chemical state of one or more layers of graphene.  

In most cases a graphene layer is in contact with substrate and strength of interaction with underlying substrate defines electronic properties of graphene. One of such interesting examples is graphene on metal surfaces~\cite{Wintterlin:2009}, where two distinct classes of weakly and strongly interacting graphene with substrate can be pointed out. For the physisorbed graphene layer, which is weakly bonded with substrate, the conical Dirac point in the graphene's electronic band structure is preserved, but small charge transfer to or from the metal substrate shifts the Fermi level ($E_F$). The electronic structure changes of free-standing graphene in this case are well explained in terms of the rigid band shift model. In the opposite case of the strong chemisorption the strong graphene-metal bonding interaction destroys the conical Dirac points leading to a more complicated picture, which can not be explained by the simple charge transfer. The strongly bonded graphene on top of Ni(111)~\cite{Nagashima:1994,Dedkov:2010} and the quasi free-standing graphene layer on 1\,ML\,Au/Ni(111)~\cite{Shikin:2000,Varykhalov:2008} can be considered as two extreme examples of chemisorbed and psysisorbed graphene, respectively.

Due to the strong Coulomb interaction of the incident electron with the image charges the EELS spectra usually show more complicated structure in comparison with the optical absorption spectra even in optical limit (momentum transfer by the incident electron $q_{\parallel}\approx0\,\AA^{-1}$) mainly due to a possibility of plasmon excitations~\cite{Raether:1980}. In the framework of dielectric formalism the observed EELS spectra can be modeled in terms of complex dielectric function $\epsilon(q,\omega)$. The presence of surface (with or without adsorbate overlayer) with different from the bulk dielectric properties additionally complicates the problem, forcing to consider the system in terms of effective dielectric function~\cite{Rocca:1995}. Following to the terminology in Ref.~\cite{Yan:2011} the influence of substrate on plasmon losses in the graphene layer may stem from two main effects: (i) the ``static'' one, due to the change of ground state wave functions of graphene as a result of hybridization with the states of the substrate and (ii) the ``dynamical'' one, which originates from the interplay of dielectric screening in graphene and the substrate. In case of nearly free-standing or weakly bonded graphene, the main effect of the substrate on the dielectric response of isolated graphene is the dynamical Coulomb interaction between induced charges in the substrate and graphene. In this case it is anticipated that the effective dielectric function of the graphene/substrate system can be constructed from the dielectric functions of isolated subsystems (graphene and substrate)~\cite{Yan:2011}. Then, in case of the free-electron like metal substrate the plasmon losses in graphene are expected to be screened nearly completely due to coupling to substrate plasmon~\cite{Yan:2011}. The transition and noble metals usually show complex rich structure which is mainly due to interband transitions~\cite{Rocca:1995,Hagelin:2005a,Hagelin:2005,Keast:2005}. In this connection it is interesting to compare EELS spectra of the graphene/TM system (TM -- transition metal) with the predictions of the theory for the free-electron like metal~\cite{Yan:2011}.

It is obvious that in case of strongly bonded graphene the ``static'' effect implying the strong band structure changes becomes very important as well as the ``dynamical'' one. In this case the changes of EELS spectra are expected to be due to the rearrangement of interband transitions and the corresponding plasmon losses. Compared to free-standing graphene, where the dipole electron transitions in the region of $\pi$ plasmon polarized in the direction perpendicular to graphene layer are prohibited by selection rules~\cite{Eberlein:2008,Marino:2004}, in the graphene coupled to a substrate such electron transitions can be very probable. Moreover, it was shown in Ref.~\cite{Nakayama:1984} that such transitions can reduce significantly the dispersion of plasmons originating from interband transitions.

Here we present the comparative study of the graphene/Ni(111) and the graphene/Au/Ni(111) systems by means of EELS spectroscopy in order to investigate the influence of the strength of interaction between graphene and substrate on the electronic properties of a graphene layer. As demonstrated, this information can be extracted from the analysis/comparison of the photoemission and loss-spectroscopy data and the obtained results are compared with those obtained on other graphene-based systems and with available band-structure calculations. 

\section{Experimental details}

In this work we have performed EELS measurements in the reflection geometry for several different energies $E_p$ of the primary electron beam at $q_{\parallel}\approx0\,\AA^{-1}$ and for different momentum transfer component $q_{\parallel}$ parallel to the surface at $E_p=100$\,eV. All EELS and valence band (VB) photoemission measurements were performed using the hemispherical electron energy analyzer SPECS PHOIBOS 150. The angle $\Phi$ between the analyzer and an electron gun was $60^\circ$ [Fig.\,1(a)]. The changing of the momentum transfer component, $q_{\parallel}$, parallel to the surface was performed by rotating the sample around the axis perpendicular to both axes of the electron gun and the energy analyzer. The $q_{\parallel}$ component in this case is given by the formula $q_{\parallel}=\sqrt{2mE_p}/{\hbar}\left(\sin \Theta_{i} - \sqrt{1-{E_{loss}}/{E_{p}}}\sin\Theta_{s}\right)$, where $E_p$ is the energy of the primary electron beam, $m$ is the mass of the free electron, $\Theta_{i}$ and $\Theta_{s}$ are the angles in the scattering plane with respect to the surface normal of the sample for incident and scattered electrons, respectively [Fig.\,1(b)]. Energy and angular resolution (an angular acceptance) of the analyzer for the primary electron energy $E_p=100 (150)$\,eV were $0.1$\,eV and $0.5^\circ$ ($q_{\parallel}=0.09 (0.11)\,\AA^{-1}$), correspondingly. The full-width-at-the-half-maximum (FWHM) of the reflection peak for the $E_p=100$ and 150\,eV was about 0.5\,eV. The base pressure during all measurements was less than $2\times10^{-10}$\,mbar. 

The Ni(111) crystal used in the present work has a disc-like shape with a thickness of about 2\,mm and radius of around 1\,cm. The cleaning procedure of the crystal was as follows. After transferring from air to the ultra-high vacuum (UHV) chamber it was firstly degassed for several hours at a temperature of $\approx600^\circ$\,C until the pressure in the UHV chamber was in the low $10^{-9}$\,mbar region. After that the Ni crystal was exposed to O$_2$ atmosphere at the pressure of $1\times10^{-6}$\,mbar and the temperature of about $700-750^{\circ}$C for about 20\,min. To remove the Ni oxide layer the crystal was subsequently cleaned by several repeated cycles of Ar$^+$-ion sputtering (the energy of the beam of 1\,keV, current  from the sample $\approx2\mu\mathrm{A}$, sputering time 30\,min), annealing ($T\approx800^{\circ}$\,C for 10\,min), and flash-annealing ($T\approx950^{\circ}$\,C for 5\,min). The quality of the crystal surface obtained after such treatment was confirmed by sharp $(1\times1)$ low-energy electron diffraction (LEED) pattern of Ni(111) surface with bright fundamental spots and a small background. 

The graphene/Ni(111) system was prepared via thermal decomposition of C$_2$H$_4$ or C$_3$H$_6$ gases on the Ni(111) single crystal surface according to the recipe described in detail in Refs.~\cite{Nagashima:1994,Dedkov:2010,Shikin:2000,Dedkov:2008a}. The graphene/1\,ML\,Au/Ni(111) system was prepared via intercalation of 5\,\AA-thick Au layer predeposited on top of the graphene layer on Ni(111) as described in Refs.~\cite{Shikin:2000,Varykhalov:2008}. The quality of the system was verified by angle-resolved photoelectron spectroscopy (ARPES) and LEED. Using LEED the sample was azimuthally rotated in a way that $q_{\parallel}$ (or scattering plane) corresponds to the $\Gamma-K$ direction of the Brillouin zone (BZ) of the graphene/Ni(111) system.

\section{Results and discussion}

\subsection{LEED and ARPES}

Prior to EELS measurements the LEED and VB ARPES in normal emission geometry characterization of samples under study was performed. In Fig.\,2 the valence band spectra of graphene/Ni(111), Au/graphene/Ni(111), and graphene/Au/Ni(111) recorded with He\,II$\alpha$ excitation energy are presented. Direct comparison of these spectra with ones in Ref.~\cite{Shikin:2000} measured at somewhat higher photon energy of $h\nu=50$\,eV allowed us to evaluate the thicknesses of the intercalated gold layer to be around 1\,ML and also make a conclusion about the completeness of the intercalation process. Indeed, after intercalation the $\pi$ states of graphene are shifted to the lower binding energy by about 2.1\,eV and their intensity is increased relative to the intensity of gold-related states indicating the transition of the strongly bonded graphene/Ni(111) system to the nearly free-standing graphene layer on Au/Ni(111)~\cite{Shikin:2000,Varykhalov:2008}. The corresponding LEED image (inset of Fig.\,2) shows the hexagonal symmetry of the graphene/Au/Ni(111) system. The noticeable increase of the background intensity in the row from initial graphene on Ni(111) (LEED is not shown) to the intercalation-like graphene/Au/Ni(111) system can be explained by the decrease of the crystal order (or by the increase of the number of defects) of the sample~\cite{Nagashima:1994,Dedkov:2008b} after deposition of gold and during subsequent intercalation of Au that results in increase of the number of scattered secondary electrons. We note that in our LEED data there is no any evidence of the $\left(2\times2\right)$ superstructure~\cite{Kang:2010} for graphene/Au/Ni(111). 

\subsection{EELS spectra for Ni(111) and graphene/Ni(111) at different $E_p$}

Figure\,3 shows the EELS spectra of clean Ni(111) and the graphene/Ni(111) sytem for different primary electron energies, $E_p$, in the range of $80-1000$\,eV and at $q_{\parallel}\approx0\,\AA^{-1}$. Both series of spectra show considerable changes for different $E_p$ indicating the interplay between surface and bulk electron energy-loss processes. Recently, EELS spectra of polycrystalline and single crystal of Ni for different $E_p$ have been reported~\cite{Hagelin:2005}. In general, spectra of clean Ni(111) presented in our work agree well with spectra from Ref.~\cite{Hagelin:2005}: the number and positions of major spectral features are almost the same. For relatively low primary electron energies ($E_p\leq150$\,eV) the main feature in the low energy-loss region of Ni spectra is a broad maximum at $6-9$\,eV with a shoulder at 3.8\,eV. According to Ref.~\cite{Hagelin:2005} this shoulder corresponds to the surface interband occupied Ni\,$3d$ to unoccupied Ni\,$4s+4p$ ($4p$ as admixture) electron transitions and to a lesser extent (since hybridization is less) to the surface interband occupied Ni\,$3d$ to unoccupied Ni\,$3d+4p$ ($4p$ as admixture) electron transitions. Authors of Ref.~\cite{Grosvenor:2006} supposed that spectral weight in this region can be contributed by the surface plasmon. The structure of the spectrum at $E_p=80$\,eV possibly reflects these transitions and surface plasmon at 5.6\,eV [6.0\,eV in~\cite{Hagelin:2005}], while at higher primary electron energies (beginning from 100\,eV) the intensity at the higher loss energies begin to dominate yielding to the broad maximum at $6-9$\,eV. The EELS intensity around 9.3\,eV [9.5\,eV in~\cite{Hagelin:2005}], which is most prominent in the spectrum collected at $E_p=500$\,eV, is considered to be due to the bulk plasmon, which is a result of oscillations of Ni\,$4s$ electrons~\cite{Hagelin:2005}. Besides, the spectral region around 7\,eV (most pronounced at $E_p=150$\,eV) comprises the transitions of Ni\,$3d$ electrons near the $E_F$ to the lower edge of Ni\,$4p$ unoccupied states~\cite{Hagelin:2005}. As the primary electron energy increases, two high energy-loss maxima at around 19\,eV and 27\,eV begin to develop indicating their bulk character. Authors of Ref.~\cite{Hagelin:2005} assign the 19\,eV [19.5\,eV in~\cite{Hagelin:2005}] maximum as a bulk plasmon due to collective oscillations of Ni $4s$ and $3d$ electrons while the 27\,eV maximum as a bulk interband electron transitions from mostly Ni\,$3d$ to Ni\,$4p$ states. 

Comparing the EELS spectra of clean Ni(111) and graphene/Ni(111) [Fig.\,3 (a) and (b), correspondingly] one can see the strong changes in the spectral shape, especially in the energy-loss region of $0-10$\,eV, where the formation of a pronounced doublet structure occurs. Taking into consideration the results of Ref.~\cite{Rosei:1984} these observations allow to conclude that new features in EELS spectra of graphene/Ni(111) can be unambiguously ascribed to a formation of graphene on top of Ni(111). By analogy with EELS measurements on graphite, in all graphene/Ni(111) EELS spectra two main energy regions can be distinguished. The $0-10$\,eV one contains the doublet structure. In graphite~\cite{Zeppenfeld:1969,Buechner:1977,Diebold:1988,Papageorgiou:2000} and free standing graphene~\cite{Eberlein:2008} the narrow intense peak is observed in this region which is well known as the $\pi$ plasmon due to excitations of the valence $\pi$ electrons. In the energy region of $10-30$\,eV the energy-loss spectrum reflects excitations of both $\pi$ and $\sigma$ valence electrons. The position of the $\pi$ plasmon in graphite at $q_{\parallel}\approx0\,\AA^{-1}$ depends on the mode of EELS experiment. EELS measurements in the reflection mode give the energy for the $\pi$ plasmon of approximately 6.5\,eV~\cite{Diebold:1988,Papageorgiou:2000}, while the measurements in the transmission mode yield the value of $\approx7$\,eV~\cite{Zeppenfeld:1969,Buechner:1977}. This discrepancy was ascribed to the higher surface sensitivity of the reflection EELS techniques and to the fact that transmission and reflection techniques probe different excitations (i.\,e. surface versus bulk excitations)~\cite{Papageorgiou:2000}. 

The 7.5\,eV energy-loss of the higher energy-loss component in the doublet structure observed in our measurements for graphene/Ni(111) at $E_p\geq150$\,eV and $q_{\parallel}\approx0$\,\AA$^{-1}$ nearly coincides with the energy position ($\approx7$\,eV) for the $\pi$ plasmon of graphite measured in transmission, while its loss energy of 6.8\,eV measured with the lower $E_p=100$\,eV is close to the energy position ($\approx6.5$\,eV) for the $\pi$ plasmon of graphite obtained in the EELS experiments in reflection. The lower energy-loss component of the doublet structure is located at around 2.8\,eV for $E_p=100$\,eV and at 3.3\,eV for $E_p\geq150$\,eV, respectively. 

Except for the doublet structure in the region of the $\pi$ plasmon the shape of the spectrum for $E_p=100$\,eV is similar to the EELS spectrum of the free standing graphene~\cite{Eberlein:2008} with reduced intensity of $\pi+\sigma$-structure relative to the one for the $\pi$ plasmon and with their maxima shifted to the lower loss energies relative to the maxima in the bulk graphite [$\approx7$\,eV and $\approx27$\,eV for the $\pi$ and $\pi+\sigma$ plasmons, correspondingly~\cite{Buechner:1977}]. Moreover, it is interesting to note here the rather good agreement in the positions of the higher loss energy component (6.8\,eV) and the $\pi+\sigma$ plasmon ($\approx19$\,eV) measured in our work at $E_p=100$\,eV and the positions of the $\pi$ (5.6\,eV) and $\pi+\sigma$ (19\,eV) plasmons measured for the bilayer graphene on SiC(0001) at $E_p=110$\,eV~\cite{Lu:2009} [for 3--4 layer graphene authors of Ref.~\cite{Lu:2009} obtain 6.3\,eV and 26\,eV, correspondingly]. These facts reflect the losses occurring mostly in the graphene overlayer and the topmost layer of Ni(111) substrate. As a confirmation of this conclusion one can consider the results of the secondary electron emission study of the graphene/Ni(111) system~\cite{Cupolillo:2010,Riccardi:2010}, where authors concluded that at $E_p=130$\,eV the electron emission is mainly determined by the graphene overlayer. The 19\,eV maximum in this case we interpret as the surface $\pi+\sigma$ plasmon like in graphite~\cite{Diebold:1988} and its 14\,eV shoulder -- the interband $\sigma\to\sigma^*$ transitions~\cite{Eberlein:2008,Marino:2004,Diebold:1988}. Also, taking into account the results of Refs.~\cite{Eberlein:2008,Marino:2004} the reduced intensity of $\pi+\sigma$-structure relative to the one of the $\pi$ plasmon can indicate the reduced macroscopic screening of the electric field of incident primary electrons by charges in the Ni substrate since the loss processes take place mainly in the graphene overlayer and in the topmost layer of Ni(111) substrate. The increasing of $E_p$ results in increasing of the role of the Ni substrate in the field screening (mainly due to interband electron transitions in it) and, consequently, in the shifting of the components of the doublet and $\pi+\sigma$-structure to higher loss energies. Moreover, the bigger shift ($\approx0.7$\,eV) and the increasing of intensity of the higher energy component relative to the shift ($\approx0.5$\,eV) and intensity of the lower energy component when going from $E_p=100$\,eV to $E_p\geq150$\,eV is associated with the higher screening by the substrate electrons for the higher energy component in the energy region of which the broad intense $6-9$\,eV maximum in the Ni dielectric response begins to develop. It should be also noted here that in the early measurements at $E_p=80$\,eV~\cite{Rosei:1984} authors obtain the energy positions for the components in the doublet structure similar to those in the present work but at $E_p=150$\,eV (3.3\,eV and 7.5\,eV). The reason for this is the different geometry of EELS experiments, so that the surface sensitivity (the screening by the substrate) in Ref.~\cite{Rosei:1984} at $E_p=80$\,eV was less (more) than in our geometry with $E_p=100$\,eV.  

Let us try to make some assumptions concerning the origin of the double peak structure in the $0-10$\,eV energy region. Rosei \textit{et al.}~\cite{Rosei:1984} assigned the higher energy component to the $\pi$ plasmon similar to the one in graphite. This plasmon in graphite is considered to be contributed by the electron transitions mainly in the region of the $M$ point of the surface BZ of graphene~\cite{Marino:2004,Buechner:1977}. From the theoretical work~\cite{Bertoni:2004} it is clear that in the graphene/Ni(111) system there is a hybridization between the C $\pi$ and Ni $3d$ valence band states resulting in the formation of the interface state at the $M$ point of BZ with binding energy of $-3.18$\,eV while the occupied $\pi$ states shifted to higher binding energy of $4.8$\,eV relative to their binding energy of $2.4$\,eV in free-standing graphene. Note that the value of $4.8$\,eV agrees well with the $4.5$\,eV binding energy position of $\pi$ valence states measured in our work by means of ARPES. To our view the increased band gap of $\approx8$\,eV between $\pi$ and C $\pi^*$-Ni $3d$ states at the $M$ point of BZ of the graphene/Ni(111) system compared to one for free-standing graphene [$4$\,eV in~\cite{Marino:2004,Bertoni:2004}] is reflected in EELS spectra of graphene/Ni(111) as a higher $\pi$ plasmon energy relative to free-standing graphene [$4.7$\,eV in~\cite{Eberlein:2008}]. However, taking into account the influence of electron transitions only around the $M$ point of BZ on the $\pi$ plasmon position seems to overestimate its expected value since the theoretical value for the interband transitions of order of 8\,eV is too high relative to the experimental position of $\leq7.5$\,eV for the $\pi$ plasmon. Decreasing of the C $2p_z$ character of the $\pi$ states due to hybridization with Ni $3d$ states at the $M$ point and increasing of the role of the transitions between C $\pi$ - Ni $3d$ occupied and unoccupied hybridized states at the $K$ point of BZ~\cite{Bertoni:2004} should shift the effective energy positions of interband transitions to sufficiently lower energies with the lower expected position for the $\pi$ plasmon. 

Considering the lower energy component in the doublet structure first as interface plasmon, authors of Ref.~\cite{Rosei:1984} made a suggestion about its origin as a result of purely kinematic effect. Opposed to a single peak of the $\pi$ plasmon observed in the EELS spectrum ($q_{\parallel}\approx0\,\AA^{-1}$) of graphite or free-standing graphene the double-peak structure in the region of $0-10$\,eV was previously observed in the graphite intercalated compounds (GIC's)~\cite{Ritsko:1979,Grunes:1983,Shung:1986}, in alkali-metal and in FeCl$_3$ intercalated SWCNTs~\cite{Liu:2003,Liu:2004}, and in the graphene/TiC(111) system~\cite{Nagashima:1993}. The latter system seems to be a similar to the studied here, the graphene/Ni(111) system, with the metal atoms (Ti or Ni) being in the direct contact with the carbon atoms of the graphene layer and because of a similarity for the $\pi$ band structure of graphene extracted from the ARPES measurements for both systems. In GIC's and intercalated SWCNTs, where the nearly pure charge transfer takes place, the lower energy-loss peak is usually associated with charge carrier plasmon due to intraband transitions of doped charge~\cite{Ritsko:1979,Grunes:1983,Shung:1986,Liu:2003,Liu:2004}, while the higher loss energy peak originates from interband transitions slightly modified with respect to ones in case of pure graphite. However, due to strong band structure changes of the graphene/Ni(111) system compared to free-standing graphene or GIC's such an interpretation for the lower energy component is obviously not appropriate.

Based on the abovementioned density-functional theory band structure calculations for the spin majority case~\cite{Bertoni:2004} and supposing that interband electron transitions occur mainly between the interface states derived from the same type of carbon atom in the unit cell of graphene (hoppings to the nearest neighbors are neglected) we interpret the lower loss energy component as transitions between $I_2$ (occupied bonding state between Ni and $fcc-top$ carbon atoms) and $I_4$ (unoccupied antibonding state between Ni and $fcc-top$ carbon atoms) interface states as denoted in Ref.~\cite{Bertoni:2004}, which have a binding energies of around $2.4$\,eV and $-0.02$\,eV, respectively and stem from the hybridization of Ni $3d$ states with $2p_z$ states of the $fcc-top$ carbon atoms. The energy difference of 2.42\,eV between these states is fairly close to the energy position of 2.8\,eV for the lower energy-loss component in the EELS spectrum, measured at $E_p=100$\,eV. To our view the possible reason for this good agreement can be explained as follows. Since the EELS spectra of graphene/Ni(111) for this primary electron energy are supposed to be most surface sensitive with small effective screening from Ni(111) substrate, we assume that these spectra reflect sufficiently the interband electron transitions in the ``isolated'' graphene/topmost layer of Ni(111) system, allowing to some extent for the direct comparison of the calculated electronic structure and the EELS spectra. 

\section{Angle-resolved EELS spectra for Ni(111) and graphene/Ni(111)}

In Fig.\,4 the EELS spectra of (a) clean Ni(111) and (b) graphene/Ni(111) for different $q_{\parallel}$ taken with an angular step of 1$^\circ$ with respect to the specular reflection along the $\Gamma-K$ direction of the graphene BZ ($E_p=100$\,eV) are presented. An additional confirmation of the interpretation of the doublet structure in the EELS spectra of graphene/Ni(111) presented in the previous section can be an observation of the nearly non-dispersive character for the lower energy component while the higher energy component has a parabolic dispersion. The absence of the dispersion of the lower energy component [Fig.\,4 (c)] points to the strong localized character of the corresponding electronic states for the transitions giving rise to the observation of this component in EELS spectra. To our view this is an indication of transitions between strongly localized interface states of the hybridized Ni\,$3d$--C\,$2p_z$ character with mainly Ni\,$3d$ contribution. As a support of strong localized character of Ni\,$3d$ states one can consider the observation of the nearly non-dispersive low energy-loss shoulder at nearly the same energy loss region in the EELS spectra of clean Ni(111) [Fig.\,4 (a)], which stems from $3d\to4s+4p,3d+4p$ interband electron transitions~\cite{Hagelin:2005}. 

For the higher energy component the main observation is its relatively strong dispersion [Fig.\,4 (c)], which is rather parabolic in the region of momenta transfer $|q_{\parallel}|\leq 1.0\,\AA^{-1}$ than linear one observed for the epitaxial graphene on SiC~\cite{Lu:2009} and for the vertically aligned single-walled carbon nanotubes (VA-SWCNTs)~\cite{Kramberger:2008}. Moreover, the bulk-like parabolic character of the dispersion in graphene/Ni(111) is reminiscent of the observations for stage-1 FeCl$_3$ GIC although the dispersion coefficient $\alpha\approx0.06$ [$E(q)=E(q=0)+\alpha({\hbar^2}/{m})q^2$] for graphene/Ni(111) obtained here is much lower than for intercalated graphite (0.68) and pure graphite (0.58)~\cite{Ritsko:1979}. Both latter dispersions were measured in the EELS experiments in transmission. Taking into account the aforementioned difference in EELS techniques in reflection and in transmission one should be cautious when comparing dispersions from the experiments measured in different modes and conditions. However, a direct comparison of dispersions for graphite from the EELS measurements in reflection~\cite{Papageorgiou:2000} with dispersion for graphene/Ni(111) measured in our work also shows that dispersion for graphite is more pronounced than for the graphene/Ni(111) system.

To our view the observation of the strongly reduced parabolic dispersion can be result of the following reasons. The linear dispersion of $\pi$ plasmon in graphene layer was explained by taking into consideration the local-field effects (LFE) in the framework of the random-phase approximation (RPA)~\cite{Kramberger:2008,Onida:2002} for the calculation of dielectric function of graphene. Inclusion of these effects in the calculation results in the mixture of transitions from the different areas in BZ of graphene and the observed linear dispersion can be roughly explained by the superposition of dispersions for the electron transitions originating in the vicinity of the $K$ and $M$ points of BZ~\cite{Kramberger:2008}. In case of free-standing graphene main contribution to the $\pi$ plasmon originates from the transitions in the area of the $M$ point of BZ~\cite{Marino:2004} as a consequence of Van Hove singularity for the locally flat bands in this point. In case of graphene/Ni(111) the theoretical calculations~\cite{Bertoni:2004} predict the strong band structure changes at the $M$ and $K$ points of BZ with formation of several interface states which are locally can be viewed as flat bands. Due to the mentioned Van Hove singularity the appearance of the interface states at the $K$ point of BZ of graphene/Ni(111) should lead to increasing the role of the electron transitions at this point in the dielectric response of the system. Because of the strong localized character of the states the corresponding transitions are reflected in the observation of non-dispersive lower-energy component in the doublet structure of EELS spectra. The mixture of the transitions in the area of the $M$ point of BZ of graphene/Ni(111), which contribute mainly in the higher-energy component in the doublet, with the nearly non-dispersive transitions at the $K$ point of BZ should result in the observed reduced parabolic dispersion for the higher-energy component ($\pi$ plasmon).

As it was mentioned in the introduction the other possible explanation for the reduced parabolic dispersion of the $\pi$ plasmon can be the possibility for the dipole electron transitions polarized in the direction perpendicular to the graphene layer on Ni(111), which are not allowed in the case of free-standing graphene~\cite{Eberlein:2008,Marino:2004}. The states for such transitions, in principle, can be formed due to hybridization with the states of the substrate. However, this explanation implies again the mixture of the different transitions polarized in and perpendicular to the graphene layer, which originate from different areas of BZ. Thus, this explanation can be regarded along with the previous explanation in terms of LFE. 

Finally, we note that the EELS spectra for high $|q_{\parallel}| > 1.0\,\AA^{-1}$ reflect rather the nearly non-dispersive energy-losses in graphene/Ni(111) due to interband electron transitions which in the framework of our interpretation may partly occur between interface states~\cite{Bertoni:2004} at the $K$ and $M$ points of the graphene BZ and dominate in this momenta transfer region.  To prove or deny our interpretation of the EELS spectra the dielectric function calculations for the graphene/Ni(111) system are necessary. 

\section{Comparison of EELS spectra for Ni(111), graphene/Ni(111) and graphene/Au/Ni(111) at $q_{\parallel}\approx0$\,\AA$^{-1}$}

Now we discuss the EELS spectra of clean Ni(111), graphene/Ni(111), 1\,ML\,Au/graphene/Ni(111), and graphene/Au/Ni(111) systems. Corresponding spectra measured at the specular reflection angle ($q_{\parallel}\approx0\,\AA^{-1}$) and primary electron energy of 150\,eV are presented in Fig.\,5. All spectra were arbitrary normalized for convenience of comparison. The shape of the EELS spectrum of the 1\,ML\,Au/graphene/Ni(111) system is very similar to one of the graphene/Ni(111) system although some important changes can be noted. Firstly, the increasing of the background takes place in the spectrum of 1\,ML\,Au/graphene/Ni(111) system. This is explained by the same reasons as the changes in the LEED images -- decreasing of the structural order due to evaporating of gold and consequently increasing of the amount of secondary electrons. Secondly, the disappearance of the low-energy loss shoulder in the $\pi+\sigma$ structure and the decrease in the intensity of the lower-energy component in the doublet structure after deposition of gold can be attributed to the fact that these losses are contributed by the losses in the Ni(111) substrate or due to interband transitions between mainly Ni-derived electronic states.

As it was mentioned above the shape of the measured EELS spectrum of graphene/Ni(111) for $q_{\parallel}\approx0\,\AA^{-1}$ is very similar to that of graphite crystal except for an additional feature at approximately $3.3$\,eV energy loss ($E_p=150$\,eV), which is characteristic of the graphene/Ni(111) interface. After intercalation of the gold layer underneath the graphene layer on Ni(111) this feature disappears and the shift of the $\pi$ plasmon to the lower loss energy is observed. Besides, the increase of the $\pi$ plasmon intensity relative to that of $\pi+\sigma$ plasmon is observed. To our view these facts indicate the transition of initial system of strongly bonded graphene on Ni(111) to a free graphene like state and correspond to a more pronounced 2D character of the losses occurring in the graphene on Au contrary to graphene on Ni(111) due to decreasing of the influence of the substrate as a result of blocking of the Ni\,$3d$--C\,$\pi$ interaction~\cite{Shikin:2000}. Indeed, blocking of the Ni\,$3d$--C\,$\pi$ interaction must result in the nearly recovery of the $\pi$ band structure of free graphene. The appearance of the intense $\pi$ plasmon can be attributed then to the $\pi$ states rearrangement resulting in the narrowing of the energy band of the possible electron transitions contributing to the intense plasmon loss.

Thus, the observed higher loss-energy positions of the $\pi$ and $\pi+\sigma$ plasmons in graphene/Au/Ni(111) compared to those in free-standing graphene can be attributed to ``dynamical'' effect of the Au/Ni(111) substrate on the graphene overlayer. Taking into consideration the EELS spectra of polycrystalline gold~\cite{Hagelin:2005a} it should be noted that the gold losses also can contribute to the loss intensity in the region of the $\pi$ plasmon in graphene/Au/Ni(111). In particular, the appearance of weak shoulder at the loss energy of $\approx3$\,eV as the primary electron energy increases (not shown here) can be ascribed to the gold losses present in this energy loss region.

\section{Conclusion}

The electronic structure of the graphene/Ni(111) and the graphene/Au/ Ni(111) systems were studied by means of angle-resolved photoelectron spectroscopy and electron energy-loss spectroscopy. A reduced dispersion of the $\pi$ plasmon excitation for the graphene/Ni(111) system was observed compared to that for pristine and intercalated graphite as well as for free-standing graphene. These observations indicate the strong band structure changes of the graphene layer on Ni(111) in comparison with free-standing graphene. It is assumed that dielectric function calculations taking into account LFE may reproduce experimentally observed spectra. We proved results of previous studies by ARPES and HREELS that intercalation of the gold layer underneath graphene on Ni(111) leads to decoupling of the electronic states of graphene and substrate and manifesting in the disappearance of the some specific EELS spectral features which are characteristic of the graphene/Ni(111) interface as well as in the shift of the $\pi$ plasmon to the lower loss energies. 

\bibliographystyle{model3-num-names}

\newpage
\textbf{Figure captions}

Figure 1: (a) The scheme and (b) geometry of EELS experiments in the present work. (For discussion of details, please see the text.)
\newline

Figure 2: The valence-band photoemission spectra measured in normal emission geometry with He\,II$\alpha$ excitation energy ($h\nu=40.8$\,eV) for graphene/Ni(111), Au/graphene/Ni(111), and graphene/Au/Ni(111). Inset shows the LEED image of the graphene/Au/Ni(111) system.
\newline

Figure 3: The EELS spectra of (a) the clean Ni(111) surface and (b) the graphene/Ni(111) system measured at different primary electron energies and at $q_{\parallel}\approx0\,\AA^{-1}$.
\newline

Figure 4: The angle-resolved EELS spectra ($E_p=100$\,eV) of (a) clean Ni(111) and (b) the graphene/Ni(111) system taken with an energy step of 1$^\circ$ with respect to the specular reflection (shown by the thick solid line) in the $\Gamma-$K direction of the first BZ of graphene. (c) The dispersion, $E_{loss}(q_{\parallel})$ of the higher (squared green symbols) and the lower (round brown symbols) energy components of the doublet structure in EELS spectra of the graphene/Ni(111) system. The parabolic fit to the dispersion of the higher energy component in the region of $|q_{\parallel}|\leq 1.0\,\AA^{-1}$ is shown by the solid blue line.
\newline

Figure 5: (a,b) The EELS spectra measured at the specular reflection angle for (1) clean Ni(111), (2) graphene/Ni(111), (3) the 1\,ML\,Au/graphene/Ni(111) and (4) the graphene/Au/Ni(111) systems with the primary electron energy of 150\,eV. The spectra were arbitrary normalized for ease of comparison. The shift to the lower loss energies of the $\pi$ plasmon of the graphene layer after intercalation of gold and the disappearing of the interface interband excitations at about 3.3\,eV are observed. (a) and (b) show wide energy-loss scan and its zoom in the energy-loss range of the $\pi$ plasmon, respectively.

\newpage
\begin{figure*}[t]
\begin{center}
\includegraphics[scale=0.6]{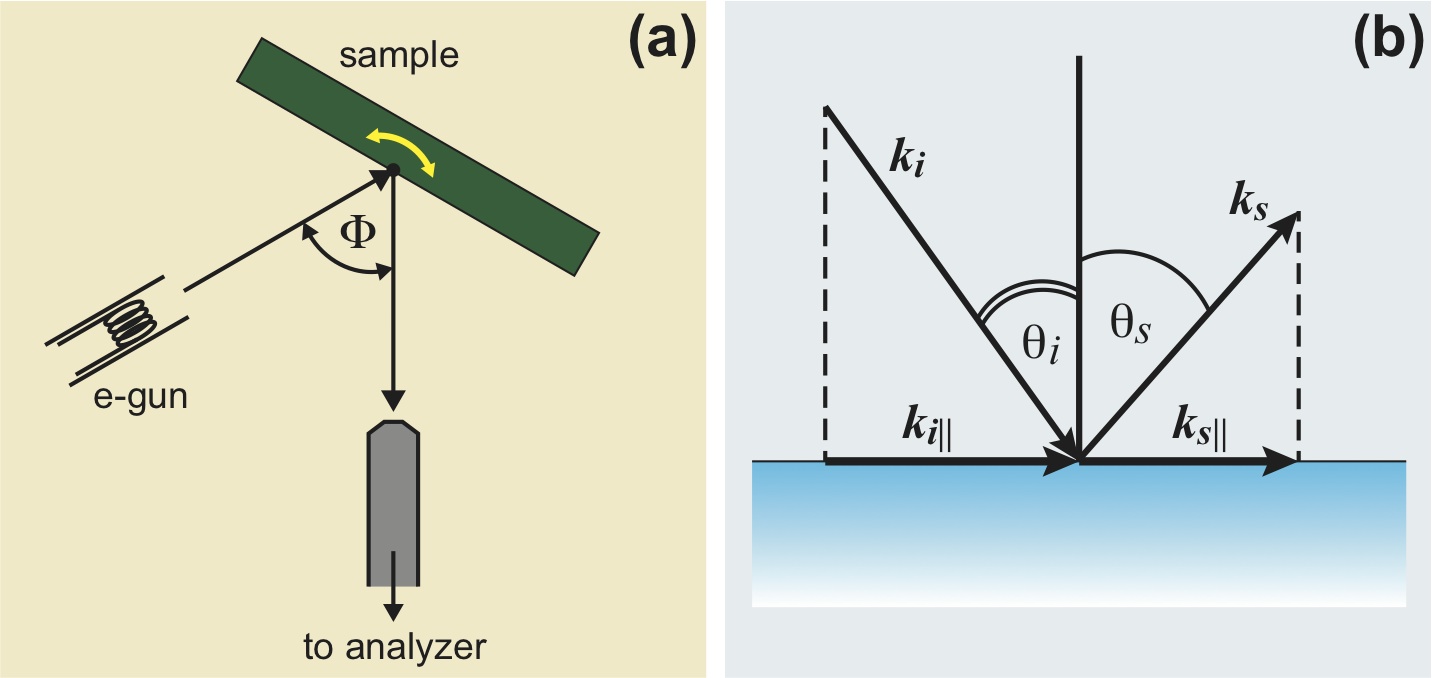}
\end{center}
\caption{(a) The scheme and (b) geometry of EELS experiments in the present work. (For discussion of details, please see the text.)}
\end{figure*}

\begin{figure}  
\begin{center}
\includegraphics[scale=0.4]{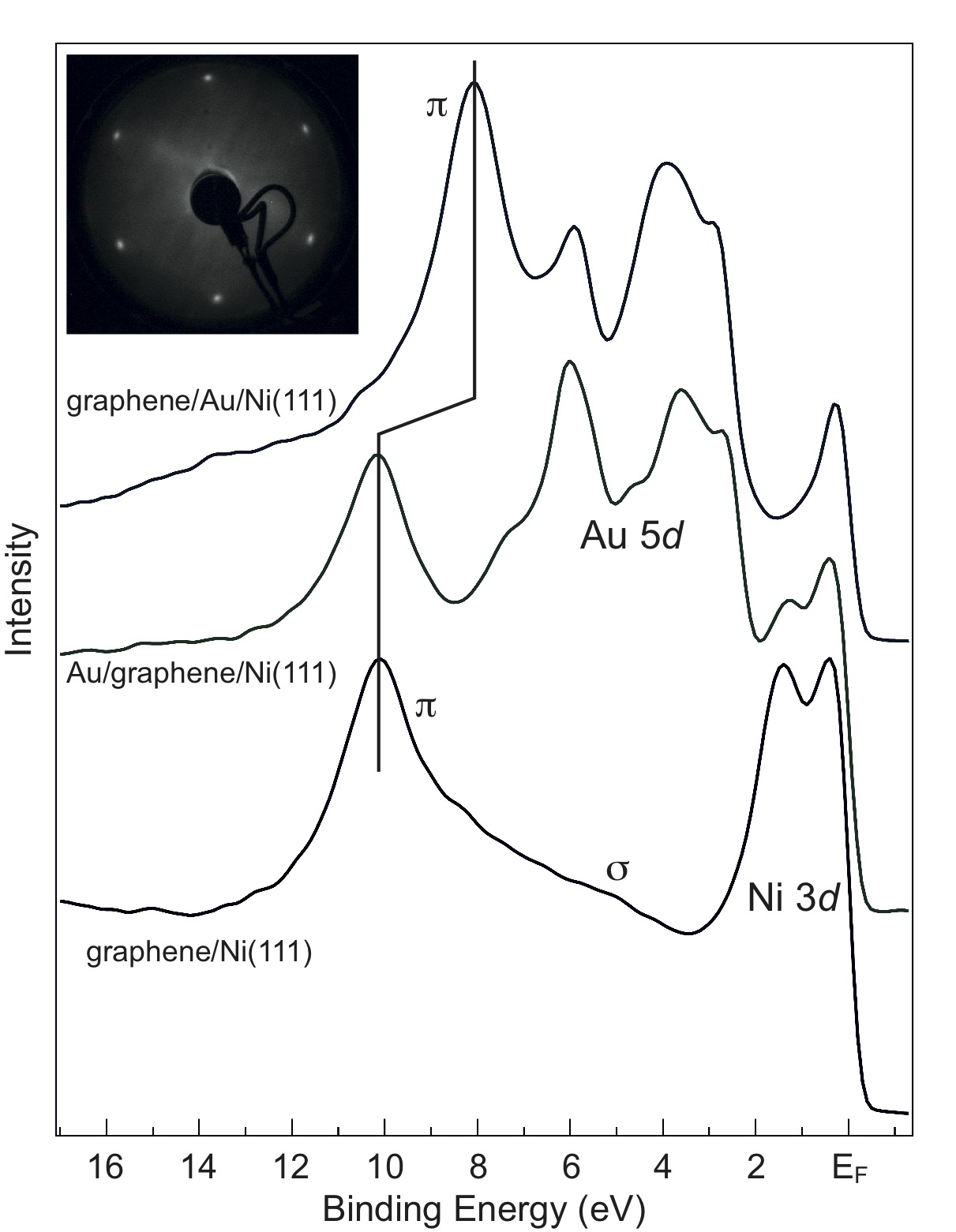}\\
\end{center}
\caption{The valence-band photoemission spectra measured in normal emission geometry with He\,II$\alpha$ excitation energy ($h\nu=40.8$\,eV) for graphene/Ni(111), Au/graphene/Ni(111), and graphene/Au/Ni(111). Inset shows the LEED image of the graphene/Au/Ni(111) system.}
\end{figure}

\begin{figure*}[t]
\begin{center}
\includegraphics[scale=0.7]{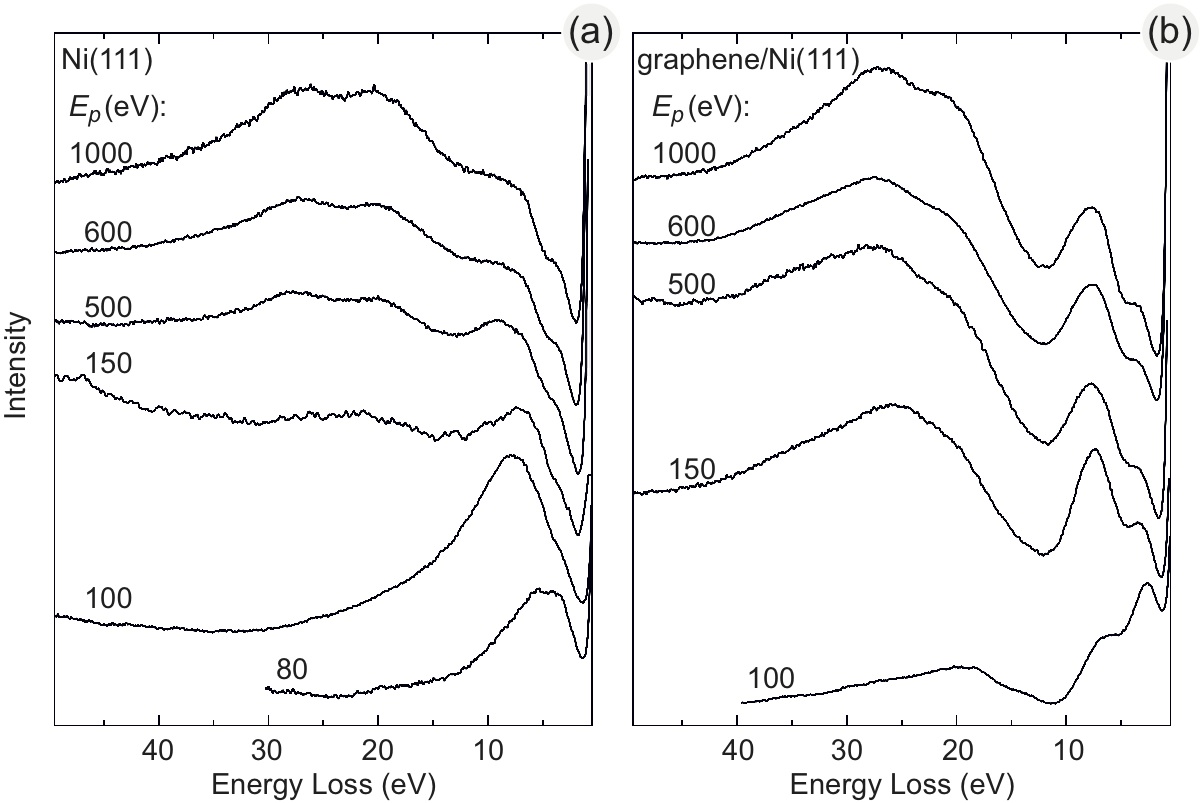}
\end{center}
\caption{The EELS spectra of (a) the clean Ni(111) surface and (b) the graphene/Ni(111) system measured at different primary electron energies and at $q_{\parallel}\approx0\,\AA^{-1}$.}
\end{figure*}

\begin{figure*}[t]
\begin{center}
\includegraphics[scale=0.45]{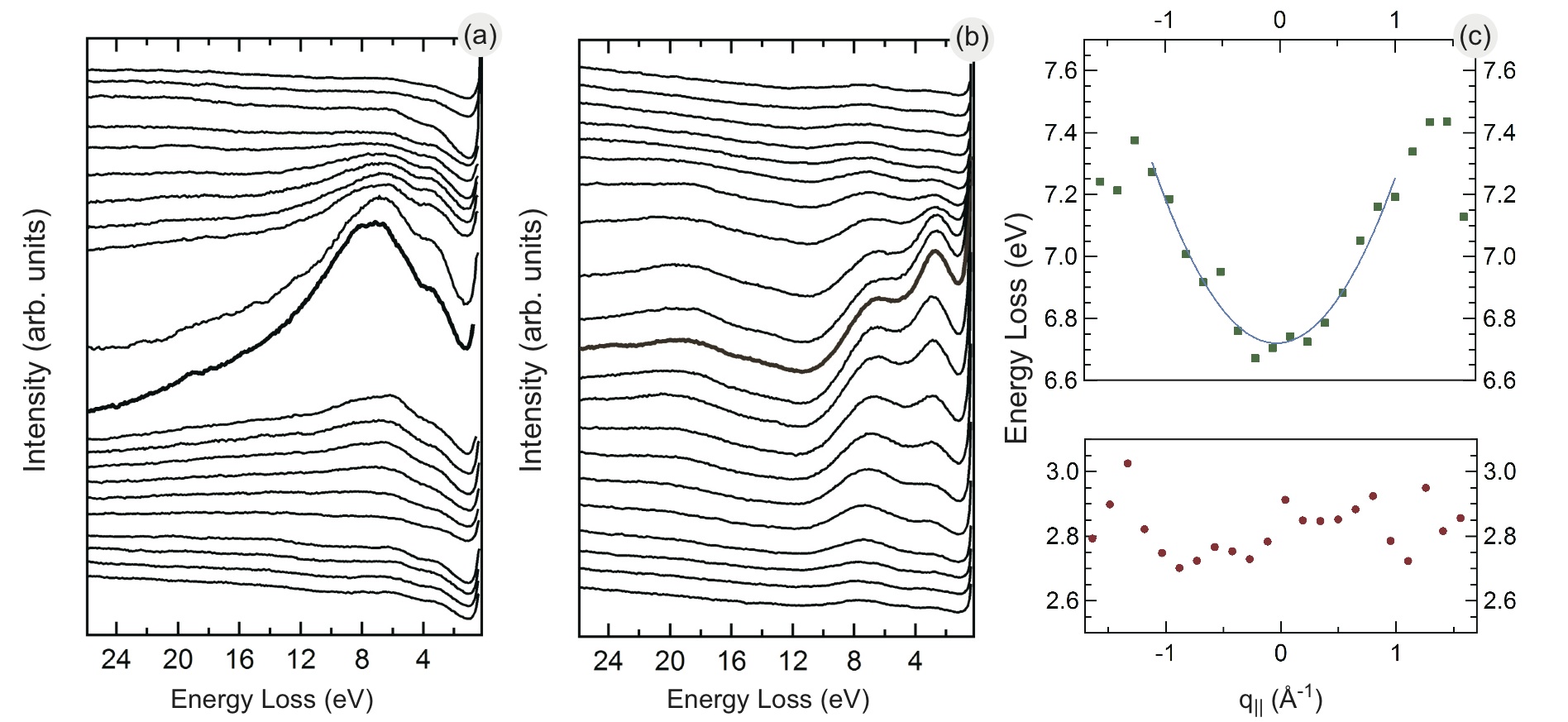}\\
\end{center}
\caption{The angle-resolved EELS spectra ($E_p=100$\,eV) of (a) clean Ni(111) and (b) the graphene/Ni(111) system taken with an energy step of 1$^\circ$ with respect to the specular reflection (shown by the thick solid line) in the $\Gamma-$K direction of the first BZ of graphene. (c) The dispersion, $E_{loss}(q_{\parallel})$ of the higher (squared green symbols) and the lower (round brown symbols) energy components of the doublet structure in EELS spectra of the graphene/Ni(111) system. The parabolic fit to the dispersion of the higher energy component in the region of $|q_{\parallel}|\leq 1.0\,\AA^{-1}$ is shown by the solid blue line.}
\end{figure*}

\begin{figure}[t]
\begin{center}
\includegraphics[scale=0.45]{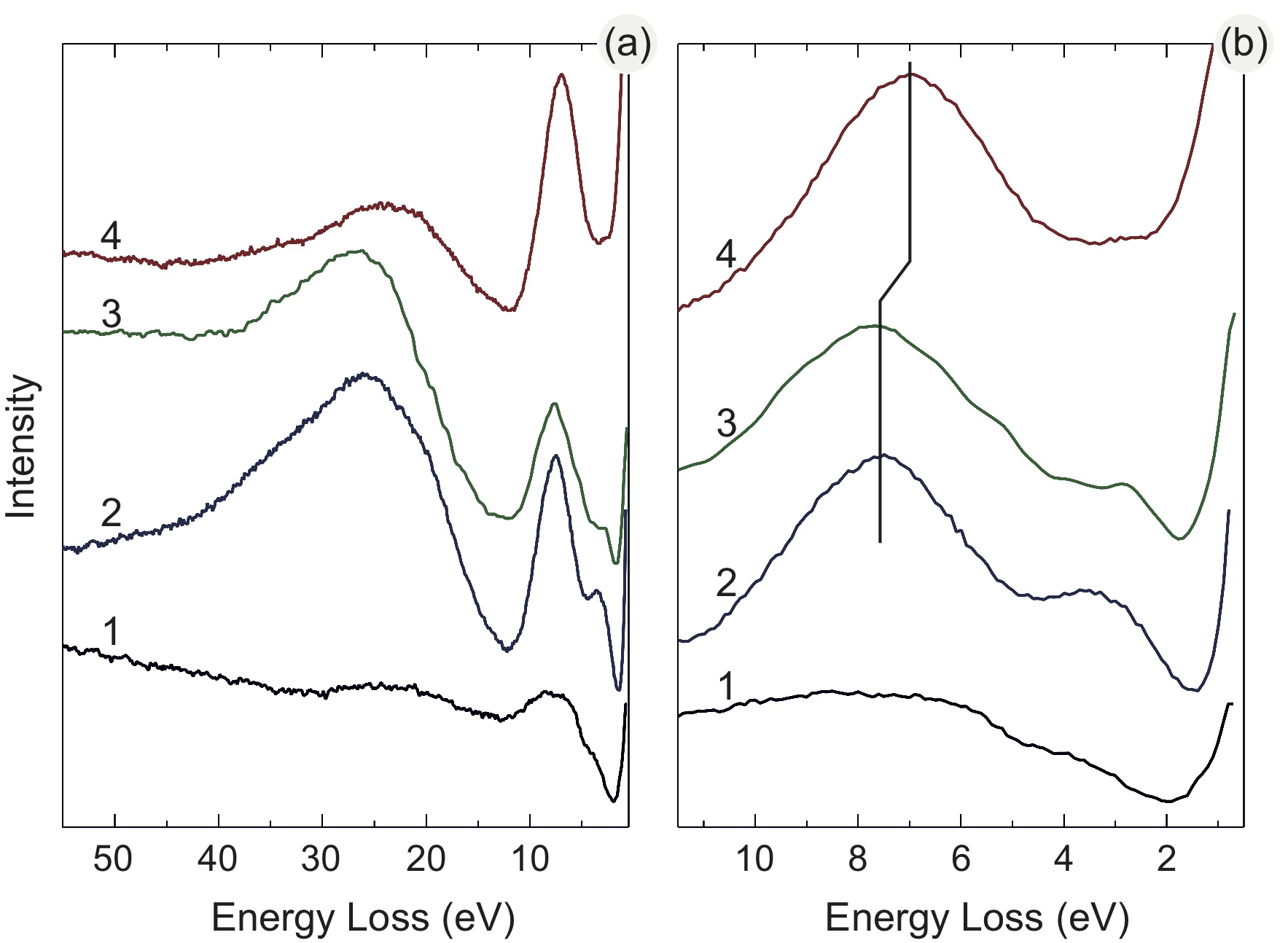}
\end{center}
\caption{(a,b) The EELS spectra measured at the specular reflection angle for (1) clean Ni(111), (2) graphene/Ni(111), (3) the 1\,ML\,Au/graphene/Ni(111) and (4) the graphene/Au/Ni(111) systems with the primary electron energy of 150\,eV. The spectra were arbitrary normalized for ease of comparison. The shift to the lower loss energies of the $\pi$ plasmon of the graphene layer after intercalation of gold and the disappearing of the interface interband excitations at about 3.3\,eV are observed. (a) and (b) show wide energy-loss scan and its zoom in the energy-loss range of the $\pi$ plasmon, respectively.}
\end{figure}

\end{document}